\documentclass[
showpacs,preprintnumbers,amsmath,amssymb]{revtex4}
\usepackage{amsmath}

\begin{document}

\tolerance=5000

\def\be{\begin{equation}}
\def\ee{\end{equation}}
\def\bea{\begin{eqnarray}}
\def\eea{\end{eqnarray}}
\def\tr{{\rm tr}\, }
\def\nn{\nonumber \\}
\def\e{{\rm e}}

\title{The future evolution and finite-time singularities in $F(R)$-gravity
unifying the inflation and cosmic acceleration}
\author{Shin'ichi Nojiri}
\affiliation{Department of Physics, Nagoya University, Nagoya 464-8602. Japan}
\author{Sergei D. Odintsov\footnote{also at Lab. Fundam. Study, Tomsk State
Pedagogical University, Tomsk}}
\affiliation{Instituci\`{o} Catalana de Recerca i Estudis Avan\c{c}ats (ICREA)
and Institut de Ciencies de l'Espai (IEEC-CSIC),
Campus UAB, Facultat de Ciencies, Torre C5-Par-2a pl, E-08193 Bellaterra
(Barcelona), Spain}

\begin{abstract}
We study the future evolution of quintessence/phantom dominated epoch in
modified $F(R)$-gravity which unifies the early-time inflation with
late-time acceleration and which is consistent with observational tests.
Using the reconstruction technique it is demonstrated that there are
models where any known (Big Rip, II, III or IV Type ) singularity may
classically occur. From another side, in Einstein frame (scalar-tensor
description) only IV Type singularity occurs. Near the singularity the
classical description breaks up, it is demonstrated that quantum
effects act against the
singularity and may prevent its appearance.
The realistic $F(R)$-gravity which is future singularity free is
proposed. We point out that additional modification of any $F(R)$-gravity
by the terms relevant at the early universe is possible, in such a way
that future singularity does not occur even classically.

\end{abstract}

\pacs{11.25.-w, 95.36.+x, 98.80.-k}

\maketitle

\section{Introduction}

The current interest to modified $F(R)$-gravity is caused by its success
as the gravitational alternative for dark energy (for a review,
see \cite{review}). The number of studies of dark energy aspects
of $F(R)$-gravity \cite{NO,FRB,FRB1} clarified its structure and possible
cosmological applications. It has been demonstrated \cite{NO} that there
appears the possibility to unify the description of the inflation and
late-time cosmic acceleration within the same theory which has standard
newtonian regime in Solar System. Hence, the whole universe expansion
history may be obtained as the cosmological solution from some modified $F(R)$-gravity.
Recently, the class of viable (non-linear, analytical) $F(R)$-gravities
was suggested in ref.\cite{HS}. Such theories and their
extensions \cite{AB,sol} do pass the local tests and successfully describe
the (almost) $\Lambda$CDM epoch. Generalizations of these theories
proposed in refs.\cite{Nojiri:2007as,Nojiri:2007cq,sergio}
keep all their nice viability properties but may simultaneously describe the
inflation, so that
whole universe expansion history: inflation, radiation/matter dominance,
dark energy epoch follows from the same viable $F(R)$-gravity consistent
with Solar System tests.

The observational data indicate that current dark universe has the
effective equation of state paremeter $w$ being very close to $-1$.
When $w=-1$ the universe passes through $\Lambda$CDM epoch. If $w$ is
slightly less
than $-1$ then we live in phantom-dominated universe and if $w$ is
slightly more than $-1$ the quintessence dark epoch occurs. In all these
cases part of (or all) energy conditions are violated.
It is known that dark energy universe with the effective phantom phase
ends up in the future, finite-time singularity \cite{BM}. From another
point, the effective quintessence universe may end up in more general
(soft) singularity.
This is also true for modified $F(R)$-gravity model\cite{review} which
leads to
dark universe with the
corresponding (phantom/quintessence) effective $w$. One can always make
the effective phantom
phase being transient, just changing the classical structure of
$F(R)$-gravity by the terms which are relevant at the very early universe
(see the corresponding investigation in \cite{abdalla}). Recently, the
appearance of future singularity in the specific example of $F(R)$-gravity
was re-obtained in refs.\cite{AB2,frlv}.

In the present work we investigate the future evolution of modified
gravity which unifies the inflation with late-time acceleration.
Using the classification of finite-time future singularities proposed in
ref.\cite{Nojiri:2005sx} we describe how all types of finite-time
singularities may
occur in modified gravity. Comparing Jordan and Einstein frames we show
that the transition of the singularity type occurs: whatever type
future singularity appears in Jordan frame ( the original $F(R)$-gravity),
in the Einstein frame (scalar-tensor description) it shows up as Type IV
singularity.
The classical mechanics analogy for future singularity is given.
As classical description breaks up near to singularity, the
quantum effects should be taken into account while approaching future
singularity. It is demonstrated how quantum effects prevent the future
singularity in modified gravity. The realistic model of non-linear
$F(R)$-gravity which unifies the inflation with late-time cosmic
acceleration without future singularity and which is consistent with local
tests is proposed. The early-time modification of the gravitational theory
may be always proposed in such a way that future singularity does not occur
while local tests are not violated.

\section{General structure of $F(R)$-gravity and trace equation}

Let us start from the general (Jordan frame) action of $F(R)$-gravity (for
a review, see \cite{review}) with matter:
\be
\label{fr1}
S_{F(R)}=\int d^4 x \sqrt{-g} \left\{\frac{F(R)}{2\kappa^2} + {\cal L}_m\right\}\ ,
\ee
The standard field equations are given by
\be
\label{JGRG13}
\frac{1}{2}g_{\mu\nu} F(R) - R_{\mu\nu} F'(R) - g_{\mu\nu} \Box F'(R)
+ \nabla_\mu \nabla_\nu F'(R) = - \frac{\kappa^2}{2}T_{(m)\mu\nu}\ .
\ee
Here $F(R)$ is a proper function of the scalar curvature $R$ and ${\cal L}_m$ is
the matter Lagrangian. In (\ref{JGRG13}), $T_{(m)\mu\nu}$ is matter energy-momentum
tensor. By separating $F(R)$ into the Einstein-Hilbert part
and modified part as
\be
\label{fr2}
F(R)=R + f(R)\ ,
\ee
the trace part of the equation of motion (\ref{JGRG13}) has the following
Klein-Gordon equation-like form:
\be
\label{Scalaron}
3\Box f'(R)=\frac{dV_{\rm eff}}{df'(R)} \equiv R+2f(R)-Rf'(R)-\kappa^2 T\ .
\ee
The above trace equation can be interpreted as an equation of motion for
the non trivial `scalaron' $f'(R)$ with the effective potential $V_{\rm eff}$.
This means that the curvature itself propagates.
The above trace equation is very convenient to study the possible
instabilities and solar system tests \cite{sol}. Note also that potential
which appears in trace equation is rather formal, it has no direct
physical meaning, like energy density or pressure.

By introducing the auxiliary field $A$, one rewrites the action
(\ref{fr1}) of the $F(R)$-gravity in the following form:
\be
\label{fr3}
S=\frac{1}{2\kappa^2}\int d^4 x \sqrt{-g} \left\{F'(A)\left(R-A\right) + F(A) + {\cal L}_m\right\}\ .
\ee
By the variation over $A$, one obtains $A=R$. Substituting $A=R$ into
the action (\ref{fr3}),
one can reproduce the action in (\ref{fr1}). Furthermore, we rescale the
metric in the following way (conformal transformation):
\be
\label{JGRG22}
g_{\mu\nu}\to \e^\sigma g_{\mu\nu}\ ,\quad \sigma = -\ln F'(A)\ .
\ee
Hence, the Einstein frame action is obtained:
\bea
\label{fr4}
S_E &=& \frac{1}{2\kappa^2}\int d^4 x \sqrt{-g} \left\{ R - \frac{3}{2}g^{\rho\sigma}
\partial_\rho \sigma \partial_\sigma \sigma - V(\sigma) + {\cal L}_m^A\right\} \ ,\nn
V(\sigma) &=& \e^\sigma G\left(\e^{-\sigma}\right)
 - \e^{2\sigma} f\left(G\left(\e^{-\sigma}\right)\right)
= \frac{A}{F'(A)} - \frac{F(A)}{F'(A)^2}
\eea
Here $G\left(\e^{-\sigma}\right)$ is given by solving the equation
$\sigma = -\ln\left( 1 + f'(A)\right)=\ln F'(A)$ as $A=g\left(\e^{-\sigma}\right)$.

In the Hu-Sawicki model \cite{HS}, when the curvature is large, $f(R)$ behaves as
\be
\label{frlv3}
f(R) \sim - 2\Lambda + \frac{\alpha}{R^n}\ .
\ee
Here $\Lambda$, $\alpha$, and $n$ are positive constants.
Then the potential $V$ is given by
\be
\label{fr5}
V = 2\Lambda + {\cal O}\left(A^{-n}\right)\ .
\ee
Therefore, the curvature infinity ($R=A\to \infty$) surely
corresponds to the finite value
of the potential.

We may consider the case that $f(R)$ behaves as
\be
\label{f1}
f(R) \sim f_n R^n\ ,
\ee
with positive $n$ and $f_n$ being a constant. The potential is found to be
\be
\label{f2}
V = \frac{n-2}{f_n A^{n-2}}\ .
\ee
Then if $2>n>0$, $V$ becomes infinite when curvature $R=A$ goes to infinity.
Therefore, in this case the singularity could not be realized easily.
In the model proposed in \cite{Nojiri:2007cq}, $f(R)$ behaves as (\ref{f1}).
Therefore, this case qualitatively similar to above behavior.

In the model \cite{Nojiri:2007as},
\be
\label{Uf5}
f(R) = - \frac{\left(R-R_0\right)^{2n+1} + R_0^{2n+1}}{f_0
+ f_1 \left\{\left(R-R_0\right)^{2n+1} + R_0^{2n+1}\right\}}
=- \frac{1}{f_1} + \frac{f_0/f_1}{f_0
+ f_1 \left\{\left(R-R_0\right)^{2n+1} + R_0^{2n+1}\right\}}\ ,
\ee
when $R$ is large, $f(R)$ behaves as
\be
\label{U1}
f(R) \sim -\frac{1}{f_1} + \frac{f_0}{f_1^2 R^{2n+1}}\ ,
\ee
which is almost the same with (\ref{frlv3}) and we obtain the expression
of $V$, which is
also similar to (\ref{fr5}),
\be
\label{U2}
V = \frac{1}{f_1} + {\cal O}\left(A^{-n}\right)\ .
\ee
In case (\ref{U1}), $1/f_1$ correspond to the effective cosmological constant
in the inflation epoch and therefore it is very large. Then since $V$
becomes positively large, the singularity could not be easily generated.
Hence, from the general structure of $F(R)$-gravity in the Einstein frame
one may deduce its behavior at large curvature.

We may include the effect of the matter with a constant equation of state (EoS)
parameter $w$. Since the matter density behaves as $\rho = \rho_0 a^{-3(1+w)}$
($\rho_0>0$) and $\rho$ has mass dimension 4, by the
scale transformation (\ref{JGRG22}), $\rho$ is transformed as
\be
\label{m0}
\tilde\rho = \e^{-2\sigma} \rho
= \rho_0 \e^{-2\sigma} {\tilde a}^{-3(1+w)} \e^{3(1+w)\sigma/2}
= \rho_0 {\tilde a}^{-3(1+w)} \e^{-(1-3w)\sigma/2}\ .
\ee
Here $\tilde a$ is the scale factor in the Einstein frame, which is given by
$\tilde a = \e^{\sigma/2} a$. Note that, the coordinates effectively
transform as $x^\mu \to {x^\mu}' = \e^{\sigma/2}$ under the scale transformation,
general mass dimension $n$ quantity $Q$ transforms as $Q\to Q' = \e^{-n\sigma/2} Q$.
Since the energy density depends on the scalar field $\sigma$, in the Einstein frame,
the potential could be shifted as
\be
\label{m}
V \to \tilde V \equiv V + C\rho_0 {\tilde a}^{-3(1+w)}\e^{- (1-3w)\sigma/2} =
V + \rho_0 {\tilde a}^{-3(1+w)}\left(F'(A)\right)^{1-3w}\ .
\ee
For the model (\ref{frlv3}), we find $F'(A)\to 1$ then the matter correction does not give
a strong effect and the singularity could be easily realized.
On the other hand, in the model (\ref{f1}), since $F'(R) \sim nf_n R^{n-1}$, the potential
$\tilde V$ blows up as $\tilde V \sim R^{(1-3w)(n-1)}$ if $w<1/3$ and $n>1$. Then the singularity
could be prevented if we include the matter.

\section{Finite-time singularities in $F(R)$-gravity}

In this section, we investigate $F(R)$-gravity models which generate
several known types of finite-time singularities.

\subsection{Big-Rip type singularity}

As the first example, we consider the case of the Big Rip singularity,
where $H$ behaves as
\be
\label{frlv10}
H=\frac{h_0}{t_0 - t}\ .
\ee
Here $h_0$ and $t_0$ are positive constants and $H$ diverges at $t=t_0$.
In order to find the $F(R)$-gravity which generates the Big Rip type singularity,
we use the method of the reconstruction, that is,
we construct $F(R)$ model realizing {\it any} given cosmology using technique
of ref.\cite{NOr}.
The general $F(R)$-gravity action with general matter is given as:
\be
\label{JGRG61}
S = \int d^4 x \sqrt{-g}\left\{F(R) + {\cal L}_{\rm matter}\right\} \ .
\ee
The action (\ref{JGRG61}) can be rewritten by using proper functions $P(\phi)$ and
$Q(\phi)$ of a scalar field $\phi$:
\be
\label{JGRG62}
S=\int d^4 x \sqrt{-g} \left\{P(\phi) R + Q(\phi) + {\cal L}_{\rm matter}\right\}\ .
\ee
Since the scalar field $\phi$ has no kinetic term, one may regard $\phi$
as an auxiliary scalar field. By the variation over $\phi$, we obtain
\be
\label{JGRG63}
0=P'(\phi)R + Q'(\phi)\ ,
\ee
which could be solved with respect to $\phi$ as $\phi=\phi(R)$. By substituting $\phi=\phi(R)$
into the action (\ref{JGRG62}), we obtain the action of $F(R)$-gravity where
\be
\label{JGRG64}
F(R) = P(\phi(R)) R + Q(\phi(R))\ .
\ee
By the variation of the action (\ref{JGRG62}) with respect to $g_{\mu\nu}$,
the equation of motion follows:
\be
\label{JGRG65}
0 = -\frac{1}{2}g_{\mu\nu}\left\{P(\phi) R + Q(\phi) \right\}
 - R_{\mu\nu} P(\phi) + \nabla_\mu \nabla_\nu P(\phi)
 - g_{\mu\nu} \nabla^2 P(\phi) + \frac{1}{2}T_{\mu\nu}
\ee
In the FRW universe, Eq.(\ref{JGRG65}) has the following form:
\bea
\label{JGRG66}
0&=&-6 H^2 P(\phi) - Q(\phi) - 6H\frac{dP(\phi(t))}{dt} + \rho \nn
0&=&\left(4\dot H + 6H^2\right)P(\phi) + Q(\phi)
+ 2\frac{d^2 P(\phi(t))}{dt} + 4H\frac{d P(\phi(t))}{dt} + p
\eea
By combining the two equations in (\ref{JGRG66}) and deleting $Q(\phi)$, we obtain
\be
\label{JGRG67}
0 = 2\frac{d^2 P(\phi(t))}{dt^2} - 2 H \frac{dP(\phi(t))}{dt} + 4\dot H P(\phi) + p + \rho\ .
\ee
Since one can redefine $\phi$ properly as $\phi=\phi(\varphi)$, we may
choose $\phi$ to be a time coordinate: $\phi=t$.
Then assuming $\rho$, $p$ could be given by the corresponding sum of
matter with a constant EoS parameters
$w_i$ and writing the scale factor $a(t)$ as $a=a_0\e^{g(t)}$ ($a_0$ : constant),
we obtain the second rank differential equation:
\be
\label{JGRG68}
0 = 2 \frac{d^2 P(\phi)}{d\phi^2} - 2 g'(\phi) \frac{dP(\phi))}{d\phi} + 4g''(\phi) P(\phi)
+ \sum_i \left(1 + w_i\right) \rho_{i0} a_0^{-3(1+w_i)} \e^{-3(1+w_i)g(\phi)} \ .
\ee
If one can solve Eq.(\ref{JGRG68}), with respect to $P(\phi)$, one can
also find the form of $Q(\phi)$ by using (\ref{JGRG66}) as
\be
\label{JGRG69}
Q(\phi) = -6 \left(g'(\phi)\right)^2 P(\phi) - 6g'(\phi) \frac{dP(\phi)}{d\phi}
+ \sum_i \rho_{i0} a_0^{-3(1+w_i)} \e^{-3(1+w_i)g(\phi)} \ .
\ee
Thus, it follows that any given cosmology can be realized by some specific
$F(R)$-gravity.

In case of (\ref{frlv10}), if we neglect the contribution from the matter,
the general solution of (\ref{JGRG68}) is given by
\be
\label{frlv11}
P(\phi) = P_+ \left(t_0 - \phi\right)^{\alpha_+}
+ P_- \left(t_0 - \phi\right)^{\alpha_-}\ ,\quad
\alpha_\pm \equiv \frac{- h_0 + 1 \pm \sqrt{h_0^2 - 10h_0 +1}}{2}\ ,
\ee
when $h_0 > 5 + 2\sqrt{6}$ or $h_0 < 5 - 2\sqrt{6}$ and
\be
\label{rlv12}
P(\phi) = \left(t_0 - \phi \right)^{-(h_0 + 1)/2}
\left( \hat A \cos \left( \left(t_0 - \phi \right) \ln \frac{ - h_0^2 + 10 h_0 -1}{2}\right)
+ \hat B \sin \left( \left(t_0 - \phi \right) \ln \frac{ - h_0^2 + 10 h_0 -1}{2}\right) \right)\ ,
\ee
when $5 + 2\sqrt{6}> h_0 > 5 - 2\sqrt{6}$.
Using (\ref{JGRG63}), (\ref{JGRG64}), and (\ref{JGRG69}), we find the form
of $F(R)$ when $R$ is large as
\be
\label{rlv13}
F(R) \propto R^{1 - \alpha_-/2}\ ,
\ee
for $h_0 > 5 + 2\sqrt{6}$ or $h_0 < 5 - 2\sqrt{6}$ case and
\be
\label{rlv14}
F(R) \propto R^{\left(h_0 + 1\right)/4} \times \left(\mbox{oscillating parts}\right)\ ,
\ee
for $5 + 2\sqrt{6}> h_0 > 5 - 2\sqrt{6}$ case.
Then $V$ (\ref{fr4}) behaves as
\be
\label{rlv15}
V \sim R^{1+\alpha_-/2}\ ,
\ee
for $h_0 > 5 + 2\sqrt{6}$ or $h_0 < 5 - 2\sqrt{6}$ case and
\be
\label{rlv16}
V \sim R^{(3 - h_0)/4} \times \left(\mbox{oscillating parts}\right)\ ,
\ee
for $5 + 2\sqrt{6}> h_0 > 5 - 2\sqrt{6}$ case. Hence, even if the curvature tends to
infinity, the potential $V$ does not always tend to infinity.
Note also that the potential is often unbounded below.

\subsection{More general singularities}

Let us investigate more general singularity
\be
\label{frlv9}
H \sim h_0 \left(t_0 - t\right)^{-\beta}\ .
\ee
Here $h_0$ and $\beta$ are constants, $h_0$ is assumed to be positive and
$t<t_0$ as it should be for the expanding universe.
Even for non-integer $\beta<0$, some derivative of
$H$ and therefore the curvature becomes singular. Since the case $\beta=1$
corresponds to
the Big Rip, which has been investigated in the last subsection, we assume $\beta\neq 1$.
Furthermore since $\beta=0$ corresponds to deSitter space, which has no singularity,
it is assumed $\beta\neq 0$.
When $\beta>1$, the scalar curvature $R$ behaves as
\be
\label{rlv16B}
R \sim 12 H^2 \sim 12h_0^2 \left( t_0 - t \right)^{-2\beta}\ .
\ee
On the other hand, when $\beta<1$, the scalar curvature $R$ behaves as
\be
\label{rlv16C}
R \sim 6\dot H \sim 6h_0 \beta \left( t_0 - t \right)^{-\beta-1}\ .
\ee

If we write
\be
\label{rlv17}
P(\phi) = \e^{-h_0 \left( t_0 - \phi \right)^{-\beta + 1}/2\left(1 - \beta\right)} S(\phi)\ ,
\ee
Eq.(\ref{JGRG68}) without matter contribution has the following Schr\"{o}dinger equation
like form:
\be
\label{rlv18}
0 = \frac{d^2 S(\phi)}{d\phi^2} + \left( \frac{5\beta h_0}{2}
\left( t_0 - \phi \right)^{-\beta - 1}
 - \frac{h_0^2}{4}\left( t_0 - \phi \right)^{-2\beta}\right) S\ .
\ee
When $\phi=t \to t_0$, in case $\beta>1$, one finds
\be
\label{rlv19}
\left|\frac{5\beta h_0}{2}
\left( t_0 - \phi \right)^{-\beta - 1}\right| \ll
\left|\frac{h_0^2}{4}\left( t_0 - \phi \right)^{-2\beta}\right|\ .
\ee
On the other hand, in case $\beta<1$, we find
\be
\label{rlv20}
\left|\frac{5\beta h_0}{2}
\left( t_0 - \phi \right)^{-\beta - 1}\right| \gg
\left|\frac{h_0^2}{4}\left( t_0 - \phi \right)^{-2\beta}\right|\ .
\ee
In any case, Eq.(\ref{rlv18}) reduces to the following form:
\be
\label{rlv21}
0 = \frac{d^2 S(\phi)}{d\phi^2} - V_0 \left( t_0 - \phi \right)^{-\alpha} S\ ,
\ee
when $\phi=t \to t_0$. Here
\be
\label{rlv22}
\begin{array}{llll}
V_0 = - \frac{5\beta h_0}{2}\ , & \alpha = \beta + 1 & \mbox{when} & \beta<1 \\
V_0 = \frac{h_0^2}{4}\ , & \alpha = 2\beta & \mbox{when} & \beta>1
\end{array}\ .
\ee
With further redefinition
\be
\label{rlv23}
y \equiv \left( t_0 - \phi \right)^{1 - \alpha/2}\ ,\quad
S = y^{\left(\alpha/4\right) \left(1 - \alpha/2\right)^{-1}} \varphi\ ,
\ee
Eq.(\ref{rlv21}) has the following form:
\be
\label{rlv24}
0 = \frac{d^2\varphi}{dy^2} - \left\{ \left(\frac{\alpha^2}{16} - \frac{\alpha}{4} \right)
\frac{1}{y^2} + \frac{4V_0}{\left(2 - \alpha\right)^2} \right\} \varphi\ .
\ee
Note that $y\to 0$ when $\phi\to t_0$ if $1 - \alpha/2 > 0$ but
$y\to \infty$ when $\phi\to t_0$ if $1 - \alpha/2 < 0$.
Then if $1 - \alpha/2 > 0$, Eq.(\ref{rlv24}) reduces to the following form
when $\phi\to t_0$:
\be
\label{rlv25}
0 = \frac{d^2\varphi}{dy^2} - \left(\frac{\alpha^2}{16} - \frac{\alpha}{4} \right)
\frac{1}{y^2} \varphi\ ,
\ee
whose general solution is given by
\be
\label{rlv26}
\varphi = A y^{\alpha/4 - 1} + B y^{-\alpha/4}\ .
\ee
Here $A$ and $B$ are constants of the integration.
On the other hand, if $1 - \alpha/2 < 0$, Eq.(\ref{rlv24}) reduces to the following form
when $\phi\to t_0$:
\be
\label{rlv27}
0 = \frac{d^2\varphi}{dy^2} + \frac{4V_0}{\left(\alpha - 2\right)^2} \varphi\ .
\ee
When $V_0 > 0$, the general solution of (\ref{rlv27}) is given by
\be
\label{rlv28}
y = \tilde A \cos \left(\omega y\right) + \tilde B \sin \left(\omega y\right) \ ,\quad
\omega \equiv \frac{2\sqrt{V_0}}{\alpha - 2}\ .
\ee
Here $\tilde A$ and $\tilde B$ are constants of the integration.
On the other hand, if $V_0 < 0$, the general solution has the following form
\be
\label{rlv28b}
y = \hat A \e^{\hat\omega y} + \hat B \e^{-\hat\omega y} \ ,\quad
\hat\omega \equiv \frac{2\sqrt{-V_0}}{\alpha - 2}\ .
\ee

 From the above analysis, one may get the asymptotic solution for $P$
when $\phi\to t_0$.
\begin{itemize}
\item {\it $\beta>1$ case:} From (\ref{rlv22}), $\alpha=2\beta>2$ and therefore
$1 - \alpha/2 = 1 - \beta < 0$, which corresponds to (\ref{rlv27}).
Since we also find $V_0>0$, the solution is given by (\ref{rlv28}).
Then by combining (\ref{frlv9}), (\ref{rlv17}),
(\ref{rlv23}), and (\ref{rlv28}), we find the following asymptotic expression of $P(\phi)$:
\bea
\label{rlv29}
P(\phi) &\sim& \e^{\left(h_0/2\left(\beta - 1\right)\right)\left(t_0 - \phi\right)^{-\beta + 1}}
\left(t_0 - \phi\right)^{\beta/2}
\left(\tilde A \cos \left(\omega \left(t_0 - \phi\right)^{-\beta + 1}\right)
+ \tilde B \sin \left(\omega \left(t_0 - \phi\right)^{-\beta + 1}\right) \right)\ ,\nn
\omega &\equiv& \frac{h_0}{2\left(\beta - 1\right)}\ .
\eea
When $\phi\to t_0$, $P(\phi)$ tends to vanish very rapidly.
By using (\ref{JGRG63}), (\ref{JGRG64}), and (\ref{JGRG69}),
$F(R)$ looks like (at large $R$)
\be
\label{rlv29B}
F(R) \propto \e^{\left(h_0/2\left(\beta - 1\right)\right)
\left(\frac{R}{12h_0}\right)^{(\beta - 1)/2\beta}}
R^{-1/4}\times\left( \mbox{oscillating part} \right)\ ,
\ee
which gives
\be
\label{rlv29C}
V \sim \e^{ - \left(h_0/2\left(\beta - 1\right)\right)
\left(\frac{R}{12h_0}\right)^{(\beta - 1)/2\beta}}
R^{9/4}\times\left( \mbox{oscillating part} \right)\ .
\ee
Near the curvature singularity $R\to \infty$, the potential
becomes small exponentially.

\item {\it $1 > \beta > 0$ case:} From (\ref{rlv22}), we find
$\alpha=\beta +1$ and therefore $1 - \alpha/2 = 1/2 - \beta/2 > 0$, which
corresponds to (\ref{rlv25}). Since
\be
\label{rlv30}
\frac{\alpha}{4} - 1 - \left( - \frac{\alpha}{4} \right)
= \beta>0\ ,
\ee
the second term in (\ref{rlv26}) dominates when $\phi\to t_0$ if $B\neq 0$.
Then by combining (\ref{frlv9}), (\ref{rlv17}),
(\ref{rlv23}), and (\ref{rlv26}), we find the following asymptotic expression of $P(\phi)$:
\be
\label{rlv31}
P(\phi) \sim B \e^{-\left(h_0/2\left(1 - \beta\right)\right)
\left(t_0 - \phi\right)^{1-\beta}}\left(t_0 - \phi\right)^{\left(\beta + 1\right)/8}\ .
\ee
Therefore,
\be
\label{rlv31C}
F(R) \sim \e^{-\left(h_0/2\left(1-\beta\right)\right)
\left( - 6\beta h_0 R \right)^{(\beta - 1)/(\beta + 1)} } R^{7/8}\ .
\ee
Eq.(\ref{rlv16C}) shows that when $\phi=t \to t_0$, $R\to \infty$ in case
$\beta > -1$ but $R\to 0$ in case $\beta < -1$.
Therefore we find the asymptotic behavior of the potential
when $R\to \infty$ as
\be
\label{rlv31D}
V \sim \e^{ \left(h_0/2\left(1-\beta\right)\right)
\left( - 6\beta h_0 R \right)^{(\beta - 1)/(\beta + 1)} } R^{9/8}
\sim R^{9/8}\ ,
\ee
which diverges in the limit of $R\to \infty$.

\item {\it $\beta<0$ case:} As in {\it $1 > \beta > 0$}
case, one gets $\alpha=\beta +1$ and therefore $1 - \alpha/2 > 0$ but since
\be
\label{rlv32}
\frac{\alpha}{4} - 1 - \left( - \frac{\alpha}{4} \right)
= \beta <0\ ,
\ee
the first term in (\ref{rlv26}) dominates when $\phi\to t_0$, if $A\neq 0$.
Hence, the asymptotic expression of $P(\phi)$ follows:
\be
\label{rlv33}
P(\phi) \sim A \e^{-\left(h_0/2\left(1 - \beta\right)\right)
\left(t_0 - \phi\right)^{1-\beta}}
\left(t_0 - \phi\right)^{- \left(\beta^2 - 6\beta + 1\right)/8}\ .
\ee
Then $F(R)$ is given by
\be
\label{rlv35}
F(R) \sim
\left( -6h_0 \beta R \right)^{\left(\beta^2 + 2\beta + 9\right)/8\left(\beta +1\right)}
\e^{-\left(h_0/2\left(1 - \beta\right)\right)
\left( -6h_0 \beta R \right)^{\left(\beta-1\right)/\left(\beta + 1\right)}} \ .
\ee
Note that $-6h_0 \beta R >0$ when $h_0, R>0$.
If $0<\beta<-1$, $R\to \infty$ when $\phi=t \to t_0$ and therefore we find
\be
\label{rlv36}
V \sim R^{2 - \left(\beta + 1\right)^2/8 - 1/\left(\beta + 1\right)}
\e^{\left(h_0/2\left(1 - \beta\right)\right)
\left( -6h_0 \beta R \right)^{\left(\beta-1\right)/\left(\beta + 1\right)}}\ ,
\ee
which diverges when $R\to \infty$.
On the other hand, if $\beta<-1$, $R\to 0$ when $\phi=t \to t_0$ and the potential has the
following form:
\be
\label{rlv36B}
V \sim R^{2 - \left(\beta + 1\right)^2/8 - 1/\left(\beta + 1\right)}\ ,
\ee
which could vanish or diverge depending on the value of $\beta$.
\end{itemize}

When $\beta>1$ in (\ref{frlv9}), $R$ behaves as in (\ref{rlv16B}), and
when $\beta<1$, the scalar curvature $R$ behaves as in (\ref{rlv16C}).
Conversely, when $R$ behaves as
\be
\label{R1}
R \sim 6\dot H \sim R_0 \left(\beta + 1\right) \left( t_0 - t \right)^{-\gamma}\ ,
\ee
if $\gamma>2$, which corresponds to $\beta = \gamma/2 >1$,
$H$ behaves as
\be
\label{R2}
H \sim \sqrt{\frac{R_0}{12}} \left( t_0 - t \right)^{-\gamma/2}\ ,
\ee
if $2>\gamma>1$, which corresponds to $1> \beta = \gamma -1 >0$,
$H$ is given by
\be
\label{R3}
H \sim \frac{R_0}{6\left( \gamma - 1\right)} \left( t_0 - t \right)^{-\gamma + 1}\ ,
\ee
and if $\gamma<1$, which corresponds to $\beta = \gamma -1 <0$, one
obtains
\be
\label{R4}
H \sim H_0 + \frac{R_0}{6\left( \gamma - 1\right)} \left( t_0 - t \right)^{-\gamma + 1}\ .
\ee
Here $H_0$ is an arbitrary constant, which is chosen to vanish in (\ref{frlv9}).
Then since $H=\dot a(t)/a(t)$, if $\gamma>2$, we find
\be
\label{R5}
a(t) \propto \exp \left( \left(\frac{2}{\gamma} -1 \right)
\sqrt{\frac{R_0}{12}} \left( t_0 - t \right)^{-\gamma/2 + 1}\right)\ ,
\ee
when $2>\gamma>1$, $a(t)$ behaves as
\be
\label{R6}
a(t) \propto \exp \left( \frac{R_0}{6\gamma\left( \gamma - 1\right)}
\left( t_0 - t \right)^{-\gamma}\right)\ ,
\ee
and if $\gamma<1$,
\be
\label{R7}
a(t) \propto \exp \left( H_0 t + \frac{R_0}{6\gamma\left( \gamma - 1\right)}
\left( t_0 - t \right)^{-\gamma}\right)\ .
\ee
In any case, there appears a sudden future singularity \cite{sudden} at
$t=t_0$.

Since the second term in (\ref{R4}) is smaller than the first term, we may
solve (\ref{JGRG68}) asymptotically as follows:
\be
\label{RR1}
P\sim P_0 \left( 1 + \frac{2h_0}{1-\beta}\left(t_0 - \phi\right)^{1-\beta}\right)\ ,
\ee
with a constant $P_0$, which gives
\be
\label{RR2}
F(R) \sim F_0 R + F_1 R^{2\beta/\left(\beta + 1\right)}\ .
\ee
Here $F_0$ and $F_1$ are constant.
When $0>\beta>-1$, we find $2\beta/\left(\beta + 1\right)<0$, which corresponds to
(\ref{frlv3}) by identifying $n = - 2\beta/\left(\beta + 1\right)$ or
$\beta = - n/(n+2)$, which could be confirmed later in (\ref{frl14}).
On the other hand when $\beta<-1$, we find $2\beta/\left(\beta + 1\right)>2$.
As we saw in (\ref{rlv13}), the $F(R)$ generates the Big Rip singularity when $R$ is large.
Then even if $R$ is small, the $F(R)$ generates a singularity where higher derivatives $H$
diverge.

Let us also investigate how effective equation of state (EoS) parameter
$w_{\rm eff}$ for (\ref{R4}) behaves when $t\sim t_0$.
In the Einstein gravity, the FRW equations are given by
\be
\label{R8}
\frac{3}{\kappa^2}H^2 = \rho\ ,\quad - \frac{1}{\kappa^2}\left(2\dot H + 3H^2\right) = p\ .
\ee
Since the EoS parameter is defined by $w=p/\rho$, even for the
$F(R)$-gravity, we may introduce the EoS parameter $w_{\rm eff}$ by
\be
\label{R9}
w_{\rm eff} = -1 - \frac{2\dot H}{3H^2}\ .
\ee
Then if $\beta = \gamma/2 >1$, it follows
\be
\label{R10}
w_{\rm eff} \sim -1 - \frac{2\beta}{3h_0} \left( t_0 - t \right)^{-1 + \beta}
\to -1 \ ,
\ee
when $t\to t_0$.
If $1> \beta = \gamma -1 >0$, we find
\be
\label{R11}
w_{\rm eff} \sim - \frac{2\beta}{3h_0} \left( t_0 - t \right)^{-1 + \beta}
\to - \infty \ .
\ee
Finally if $\beta = \gamma -1 <0$, one gets the expression (\ref{R11})
when $H_0=0$. When $H_0\neq 0$, on the other hand, we obtain
\be
\label{R12}
w_{\rm eff} \sim -1 - \frac{2\beta h_0}{3H_0^2}
\left( t_0 - t \right)^{-1 - \beta}\ ,
\ee
when $H_0$ vanishes. Then if $-1< \beta <0$, $w_{\rm eff}\to +\infty$
when $t\to t_0$. On the other hand, if $\beta<-1$, $w_{\rm eff}\to -1$.

Eq.(\ref{R8}) also shows that, even for $F(R)$-gravity, we may define the
effective energy
density $\rho_{\rm eff}$ and the effective pressure $p_{\rm eff}$ by
\be
\label{R12B}
\rho_{\rm eff}\equiv \frac{3}{\kappa^2}H^2 ,\quad
p_{\rm eff} \equiv - \frac{1}{\kappa^2}\left(2\dot H + 3H^2\right)\ .
\ee
We now assume $H$ behaves as (\ref{frlv9}).
Then if $\beta> 1$, when $t\to t_0$,
$a\sim \exp( h_0\left( t_0 - t \right)^{1-\beta}/\left( \beta -1 \right) )
\to \infty$ and
$\rho_{\rm eff} ,\, |p_{\rm eff}| \to \infty$.
If $\beta=1$, we find
$a\sim \left(t_0 - t\right)^{-h_0} \to \infty$ and
$\rho_{\rm eff} ,\, |p_{\rm eff}| \to \infty$.
If $0<\beta<1$, $a$ goes to a constant but $\rho ,\, |p| \to \infty$.
If $-1<\beta<0$, we find $a$ and $\rho$ vanishes
but $|p_{\rm eff}| \to \infty$.
When $\beta<0$, instead of (\ref{frlv9}), as in (\ref{R3}), one may assume
\be
\label{R13}
H \sim H_0 + h_0 \left(t_0 - t\right)^{-\beta}\ .
\ee
Hence, $\rho_{\rm eff}$ has a finite value $3H_0^2/\kappa^2$
in the limit $t\to t_0$ when $-1<\beta<0$.
If $\beta<-1$ but $\beta$ is not any integer, $a$ is finite and $\rho_{\rm eff}$ and $p_{\rm eff}$
vanishes if $H_0=0$ or $\rho_{\rm eff}$ and $p_{\rm eff}$ are finite if $H_0\neq 0$
but higher derivatives of $H$ diverge.

In \cite{Nojiri:2005sx}, there was suggested the classification of the
finite-time singularities in the following way:
\begin{itemize}
\item Type I (``Big Rip'') : For $t \to t_s$, $a \to \infty$,
$\rho \to \infty$ and $|p| \to \infty$. This also includes the case of
$\rho$, $p$ being finite at $t_s$.
\item Type II (``sudden'') : For $t \to t_s$, $a \to a_s$,
$\rho \to \rho_s$ and $|p| \to \infty$
\item Type III : For $t \to t_s$, $a \to a_s$,
$\rho \to \infty$ and $|p| \to \infty$
\item Type IV : For $t \to t_s$, $a \to a_s$,
$\rho \to 0$, $|p| \to 0$ and higher derivatives of $H$ diverge.
 This also includes the case when $p$ ($\rho$) or both of them tend to
some finite values while higher derivatives of $H$ diverge.
\end{itemize}
Here $t_s$, $a_s$ and $\rho_s$ are constants with $a_s\neq 0$. We now identify $t_s$ with $t_0$.
The Type I corresponds to $\beta>1$ or $\beta=1$ case, Type II to $-1<\beta<0$ case, Type III
to $0<\beta<1$ case, and Type IV to $\beta<-1$ but $\beta$ is not any integer case.
Thus, we have constructed $F(R)$-gravity examples which show any type
of above finite-time singularity. This is natural because it is known that
modified gravity may lead to the effective phantom/quintessence phase
\cite{review} while the phantom/quintessence dominated universe may end up
with finite-time singularity.

The reconstruction method also tells that there appear Type I singularity for $F(R) = R + \alpha R^n$
with $n>2$ and Type III singularity for $F(R) = R - \beta R^{-n}$ with $n>0$.

\subsection{Classical mechanics analogy}

Let us now consider the analogy with the classical mechanics in the
description of finite-time singularities in modified gravity.

We start with the trace equation:
\be
\label{frlv1}
2F(R) - R F'(R) - 3 \Box F'(R) = - \frac{\kappa^2}{2} T\ .
\ee
By writing $F(R)$ as $F(R)=R+f(R)$, Eq.(\ref{frlv1}) could be rewritten as
\be
\label{frlv2}
R + 2f(R) - Rf'(R) - 3\Box f'(R) = - \frac{\kappa^2}{2} T\ .
\ee
When the curvature is large, we assume
$f(R)$ behaves as (\ref{frlv3}). Then Eq.(\ref{frlv2}) reduces to
\be
\label{frlv4}
R + 3\alpha \Box \left(R^{-n-1}\right) \sim 0\ .
\ee
If
\be
\label{frlv5}
\chi \equiv R^{-n - 1}\ ,
\ee
and the FRW metric with flat spatial part is chosen, Eq.(\ref{frlv4})
has the following form:
\be
\label{frlv6}
\ddot \chi + 3H \dot \chi = \frac{1}{3\alpha}\chi^{- 1/(n+1)}\ .
\ee
Note $\chi=0$ corresponds to the curvature singularity $R\to\infty$.
Neglecting the second term in the l.h.s. of (\ref{frlv6}), the equation
looks very similar
to the equation of the motion in the classical system with repulsive force
$(1/3\alpha)\chi^{- 1/(n+1)}$.

First we consider the classical equation of motion:
\be
\label{frlv7}
\ddot x = \frac{1}{3\alpha}x^{- 1/(n+1)}\ .
\ee
For Eq.(\ref{frlv7}), one gets an exact solution:
\be
\label{frlv8}
x= C \left(t_0 - t\right)^{2(n+1)/(n+2)}\ .
\ee
Here $C$ and $t_0$ are constants. Note $2>2(n+1)/(n+2)>1$.
Then $x$ vanishes in a finite time $t=t_0$, which corresponds to the curvature singularity
in (\ref{frlv7}).

We now investigate the asymptotic solution when the curvature is large, that is, $\chi$
is small. As there is a curvature singularity, one may assume
(\ref{frlv9}).
Since
\be
\label{frlv9B}
R\sim \frac{12h_0^2}{\left(t_0 - t\right)^{2\beta}}
 - \frac{6h_0 \beta }{\left(t_0 - t\right)^{\beta+1}}\ ,
\ee
$R$ diverges when $\beta>-1$ but $\beta=0$.

We now consider three cases: 1) $\beta=1$, 2) $\beta>1$, 3) $0<\beta<1$,
and 4) $0>\beta>-1$.
\begin{itemize}
\item In case 1) $\beta=1$, since
\be
\label{frl1}
R\sim \frac{12h_0^2 + 6h_0}{\left(t_0 - t\right)^{2}}\ ,
\ee
and therefore, from (\ref{frlv5}), we find
\be
\label{frl2}
\chi \sim \left(t_0 - t\right)^{2(n+1)}\ ,
\ee
and the l.h.s. of (\ref{frlv6}) behaves as
\be
\label{frl3}
\ddot \chi + 3H \dot \chi \sim \left(t_0 - t\right)^{2n}\ ,
\ee
but the r.h.s. behaves as
\be
\label{frl4}
\frac{1}{3\alpha}\chi^{- 1/(n+1)} \sim \left(t_0 - t\right)^{-2}\ ,
\ee
which is inconsistent since the powers of the both sides do not coincides
with each other. Therefore, $\beta\neq 1$.
\item In case 2) $\beta>1$, we find
\be
\label{frl5}
R = 12H^2 + 6\dot H \sim 12 H^2 \sim \left(t_0 - t\right)^{-2\beta}\ ,
\ee
and therefore
\be
\label{frl6}
\chi \sim \left(t_0 - t\right)^{2\beta(n+1)}\ .
\ee
In the l.h.s. of (\ref{frlv6}), the second term
dominates and the l.h.s. behaves as
\be
\label{frl7}
\ddot \chi + 3H \dot \chi \sim 3H \dot \chi \sim
\left(t_0 - t\right)^{\beta(2n+1) -1}\ .
\ee
On the other hand, the r.h.s. behaves as
\be
\label{frl8}
\frac{1}{3\alpha}\chi^{- 1/(n+1)} \sim \left(t_0 - t\right)^{-2\beta}\ .
\ee
Then by comparing the powers of the both sides, one gets
\be
\label{frl9}
\beta(2n+1) -1 = -2\beta\ ,
\ee
which gives $\beta = 1/(2n+3)$ but this conflicts with the assumption
$\beta>1$.
\item In case 3) $0<\beta<1$ or case 4) $0>\beta>-1$, we find
\be
\label{frl10}
R = 12H^2 + 6\dot H \sim \dot H \sim \left(t_0 - t\right)^{-\beta -1}\ ,
\ee
and therefore
\be
\label{frl11}
\chi \sim \left(t_0 - t\right)^{(\beta+1)(n+1)}\ .
\ee
Then in the l.h.s. of (\ref{frlv6}),
the first term dominates and the l.h.s. behaves as
\be
\label{frl12}
\ddot \chi + 3H \dot \chi \sim \ddot \chi
\sim \left(t_0 - t\right)^{\beta(n+1) + n -1}\ .
\ee
On the other hand, the r.h.s. behaves as
\be
\label{frl13}
\frac{1}{3\alpha}\chi^{- 1/(n+1)} \sim
\left(t_0 - t\right)^{-\beta -1}\ .
\ee
Then by comparing the powers of the left-hand side and the right-hand side,
the consistency tells
\be
\label{frl14}
\beta(n+1) + n - 1 = -\beta - 1\ \mbox{or}\ \beta = -n/(n+2)\ .
\ee
This conflicts with the case 3) $0<\beta<1$ but
consistent with the case 4) $0>\beta>-1$, which could correspond to the case in \cite{frlv}
or to (\ref{rlv31D}).
In fact, by substituting (\ref{frl14}) into (\ref{frl11}), we get
\be
\label{frl14B}
\chi \sim \left(t_0 - t\right)^{2(n+1)/(n+2)}\ .
\ee
which corresponds to (\ref{frlv8}).
Since $0>\beta>-1$, this singularity corresponds to Type III in \cite{Nojiri:2005sx}.
\end{itemize}
Then we found the curvature singularity really appears in the
Hu-Sawicki model in a finite time as pointed out in \cite{frlv}.

We now briefly comment what could happen near the singularity in the
Einstein frame (\ref{fr4}).
Since $\sigma \sim 0$ and therefore $\e^\sigma \sim 1$ in the scale transformation (\ref{JGRG22}),
we may identify the time coordinate in the Einstein frame and the time coordinate in the original
$F(R)$ frame. Since (\ref{JGRG22}) shows also
\be
\label{st1}
\sigma \sim \frac{n\alpha}{A^{n+1}} = \frac{n\alpha}{R^{n+1}}\ ,
\ee
by using (\ref{frl10}) and (\ref{frl14}), one obtains
\be
\label{st2}
\sigma \sim \frac{n\alpha}{(6h_0)^{n+1}}\left(t_0 - t\right)^{2(n+1)/(n+2)}\ ,
\ee
and therefore
\be
\label{st2B}
\dot\sigma \sim \frac{2n(n+1)\alpha}{(6h_0)^{n+1}(n+2)}\left(t_0 - t\right)^{n/(n+2)}\ .
\ee
On the other hand, the potential $V(\sigma)$ in (\ref{fr4}) has the following form:
\be
\label{st3}
V(\sigma) \sim 2\Lambda - \frac{\left(n+1\right)\alpha}{A^n}
\sim 2\Lambda - \frac{\left(n+1\right)\alpha}{(6h_0)^n} \left(t_0 - t\right)^{2n/(n+2)}\ .
\ee
Then in the Einstein frame, the energy density $\rho_\sigma$ and pressure $p_\sigma$,
which are given by
\be
\label{st4}
\rho_\sigma = \frac{3}{2}{\dot\sigma}^2 + V(\sigma)\ ,\quad
p_\sigma = \frac{3}{2}{\dot\sigma}^2 - V(\sigma)\ ,
\ee
and therefore the curvature in the Einstein frame $R_E = - \left(\kappa^2/2\right)T
= \left(\kappa^2/2\right)\left(\rho - 3p\right)$ are finite even if $t\to t_0$ since
${\dot\sigma}^2 \sim V - 2\Lambda \sim \left(t_0 - t\right)^{2n/(n+2)}$ vanishes.
As the exponent $2n/(n+2)$ is fractional in general, however, the
higher derivative of $H$ could diverge. Then in the Einstein frame, the
finite-time singularity appears as Type IV singularity.
Thus, we discovered the possibility of change of finite-time singularity nature
with the change of frame in modified gravity. This is not strange. Indeed,
it was shown some time ago \cite{fabio} that Big Rip singularity which
appears in
scalar-tensor theory qualitatively changes its form in the equivalent
Jordan frame. Exactly in the time when scalar-tensor theory becomes
singular, the mathematically-equivalent $F(R)$-gravity (which is not
completely physically equivalent to scalar-tensor theory \cite{salva})
becomes complex \cite{fabio}.


The important remark is in order.
The general (Jordan frame) action of $F(R)$-gravity with matter in (\ref{fr1}) 
can be rewritten in the Einstein frame action (\ref{fr4}). 
One should note that the sign in front of the kinetic term in  (\ref{fr4}) is 
always canonical, and therefore
there never appears  phantom in the transformed Einstein frame.
Note that this transformation is exact one.

In the Einstein frame, the FRW equations have the following form:
\be
\label{SS1}
\frac{3}{\kappa^2} H^2 = \rho_\sigma + \rho\ ,\quad
- \frac{1}{\kappa^2} \left( 3H^2 + 2\dot H \right) = p_\sigma + p\ .
\ee
Here $\rho$ and $p$ express the contribution from the matter and $\rho_\sigma$ and $p_\sigma$ 
are defined by (\ref{st4}). 
If we neglect the contribution from the matter, the EoS parameter is given by
$w=p_\sigma/\rho_\sigma$ and therefore $w>-1$ if we assume $V(\sigma)>0$, which corresponds to
the fact that $\sigma$ cannot be phantom in the Einstein frame.

When the Hubble rate $H$ is given by
\be
\label{SS3}
H \sim h_0 \left( t_0 - t \right)^{-\beta}\ ,
\ee
the effective EoS parameter looks like
\be
\label{SS4}
w_{\rm eff}\equiv -1 - \frac{2\dot H}{3H^2} \sim - 1 - \frac{2\beta}{3h_0}
\left( t_0 - t \right)^{\beta - 1} \ .
\ee
Since we are considering the period $t<t_0$ and the universe is expanding, we
assume $h_0>0$. Then $w_{\rm eff}$ can be greater than $-1$ when $\beta<0$, which corresponds to the
Type II ($0>\beta>-1$) or Type IV ($\beta<-1$).
Since the EoS parameter is greater than $-1$ in the Einstein frame,
Type I (Big Rip type $\beta\geq 1$) and III ($1>\beta>0$) singularities
are not allowed (the phantom absence) but there can occur Type II or IV singularity.

Even if there is no singularity in Einstein frame, there could appear a
finite time singularity in the Jordan frame.
We may consider the following action in the Einstein frame (\ref{fr4}), 
where the potential is given by
\be
\label{SS5}
V(\sigma) = V_0 \e^{\sigma \sqrt{\frac{3}{2h_0}}}\ .
\ee
Here $V_0$ and $h_0$ are positive constants.
Then by assuming the FRW universe in the Einstein frame
\be
\label{SS5b}
ds_{\rm E}^2 = - dt_{\rm E}^2 + a_{\rm E} \left(t_{\rm E}\right) ^2
\sum_{i=1,2,3} \left(dx^i \right)^2 \ ,
\ee
one obtains a solution:
\be
\label{SS6}
H=\frac{h_0}{t_{\rm E}}\quad \left(a_{\rm E}\left(t_{\rm E}\right) \propto
t_{\rm E}^{h_0} \right)\ ,\quad
\sigma = \sqrt{\frac{2h_0}{3}}\ln \frac{t_{\rm E}}{t_{{\rm E}\,0}}\ .
\ee
Then there is no singularity although $a$ becomes infinite in the limit of $t_{\rm E}\to +\infty$.
Since the metric tensor $g_{{\rm E}\,\mu\nu}$ in the Einstein frame
is related with the tensor $g_{{\rm J}\mu\nu}$ in the Jordan frame by
$g_{{\rm E}\,\mu\nu}=\e^{-\sigma}g_{{\rm J}\mu\nu}$, the metric
in the Jordan frame is given by
\be
\label{SS6b}
ds_{\rm J}^2 = \e^{-\sigma} \left(- dt_{\rm E}^2
+ a_{\rm E} \left(t_{\rm E}\right) ^2 \sum_{i=1,2,3} \left(dx^i \right)^2 \right)\ .
\ee
Since the FRW universe in the Jordan frame is given by
\be
\label{SS6c}
ds_{\rm J}^2 = - dt_{\rm J}^2 + a_{\rm J} \left(t_{\rm J}\right) ^2
\sum_{i=1,2,3} \left(dx^i \right)^2 \ ,
\ee
by comparing (\ref{SS6b}) with (\ref{SS6c}), the time coordinate $t_{\rm J}$
in the Jordan frame is given by
\be
\label{SS8}
t_{\rm J} - t_{{\rm J}\,0} = \pm \int dt_{\rm E} \e^{-\sigma(t_{\rm E})/2}
= \pm \frac{t_{{\rm E}\,0}}{1 - \frac{h_0}{6}} \left(\frac{t_{\rm E}}{t_{{\rm E}\,0}}
\right)^{1 - \frac{h_0}{6}}\ .
\ee
One now choose the sign $\pm$ so that the directions of the time in the
Einstein frame and the Jordan frame could not be changed.
If $1 - \frac{h_0}{6}<0$, i.e., $h_0>6$, the limit $t\to \infty$ corresponds
to $t_{\rm J} \to t_{{\rm J}\,0}$.
Then we find
\be
\label{SS10}
a_{\rm J}\left(t_{\rm J}\right) = \e^{-\sigma/2} a_{\rm E}\left(t_{\rm E}\right)
\propto t_{\rm E}^{h_0 - \frac{h_0}{6}}
\propto \left(t_{{\rm J}\,0} - t_{\rm J}\right)^{\frac{h_0 - \frac{h_0}{6}}{1 - \frac{h_0}{6}}}\ .
\ee
Hence, if $h_0>6$, $a_{\rm J}\left(t_{\rm J}\right)$ diverges at $t_{\rm
J}\to t_{{\rm J}\,0}$, that is there appears a finite singularity even
if there is no singularity in the Einstein frame.

\subsection{Quantum effects near finite-time singularity}

Near the future singularity at $t=t_0$, the curvature becomes large in
general. As a result, near the singularity the quantum fields/quantum
gravity effects become very important again.
All classical considerations are not valid, all speculations about
future cosmic doomsday cannot restrict the classical theory structure
because
quantum effects can stop (or shift) the future singularity \cite{sudden1}.
Moreover, the quantum corrections usually contain the curvature powers,
which become important near the singularity. Hence, any claim about the
appearance of the effective phantom/quintessence phase in modified gravity
which subsequently enters the future singularity is not justified without
quantum effects account near to singularity.
One may include the massless quantum effects by taking
into account the conformal anomaly contribution as back-reaction near the
singularity \cite{sudden1}.
The conformal anomaly $T_A$ has the following well-known form:
\be
\label{OVII}
T_A=b\left(F+ \frac{2}{3}\Box R\right) + b' G + b''\Box R\ ,
\ee
where $F$ is the square of 4d Weyl tensor, $G$ is Gauss-Bonnet invariant,
which are given by
\bea
\label{GF}
F&=&\frac{1}{3}R^2 -2 R_{\mu\nu}R^{\mu\nu}+ R_{\mu\nu\rho\sigma}R^{\mu\nu\rho\sigma}\ , \nn
G&=&R^2 -4 R_{\mu\nu}R^{\mu\nu}+ R_{\mu\nu\rho\sigma}R^{\mu\nu\rho\sigma}\ .
\eea
In general, with $N$ scalar, $N_{1/2}$ spinor, $N_1$ vector fields, $N_2$ ($=0$ or $1$)
gravitons and $N_{\rm HD}$ higher derivative conformal scalars, $b$ and
$b'$ are given by
\bea
\label{bs}
&& b= \frac{N +6N_{1/2}+12N_1 + 611 N_2 - 8N_{\rm HD}}{120(4\pi)^2}\ ,\nn
&& b'=- \frac{N+11N_{1/2}+62N_1 + 1411 N_2 -28 N_{\rm HD}}{360(4\pi)^2}\ .
\eea
As is seen $b>0$ and $b'<0$ for the usual matter except the higher
derivative conformal scalars.
Notice that $b''$ can be shifted by the finite renormalization of the
local counterterm $R^2$, so $b''$ can be an arbitrary coefficient.

By including the trace anomaly, Eq.(\ref{frlv2}) is modified as
\be
\label{CA1}
R + 2f(R) - Rf'(R) - 3\Box f'(R) = - \frac{\kappa^2}{2} \left(T + T_A \right)\ .
\ee
For FRW universe, we find
\be
\label{CA2}
F=0\ ,\quad G=24\left(\dot H H^2 + H^4\right)\ .
\ee
We now assume $H$ behaves as (\ref{frlv9}) and neglect the contribution from matter by
putting $T=0$.
Then in case of the Hu-Sawicki model, we assume $f(R)$ behaves as (\ref{frlv3})
and Eq.(\ref{CA1}) becomes
\be
\label{CA3}
R + 3\Box \left(\frac{\alpha n}{R^{n+1}}\right) = - \frac{\kappa^2}{2} T_A\ .
\ee
First we consider the case that $2b/3 + b''=0$ and therefore $T_A=G$.
If $-1<\beta<0$, $R$ behaves as $\left(t_0 - t \right)^{-\beta - 1}$ and
$G$ behaves as $G\sim 24 \dot H H^2 \sim \left(t_0 - t \right)^{-3\beta -1}$.
Since $-3\beta -1 > -\beta - 1$ when $\beta<0$, $T_A$ can be neglected compared with
$R$. Here, $\Box \left(\alpha n/R^{n+1}\right) \sim
\left(t_0 - t \right)^{\left(\beta + 1\right)\left(n+1\right) -2}$.
Then by comparing the r.h.s. with the l.h.s. in (\ref{CA3}), one reobtains
(\ref{frl14}).
Therefore, the curvature singularity appears in a finite time and the
quantum correction
does not prevent the singularity when $2b/3 + b''=0$.
We may, however, consider the case $2b/3 + b''\neq 0$. In this case, $T_A$
behaves as
\be
\label{CA4}
T_A \sim \left(\frac{2}{3}b + b'' \right) \Box R \sim \left(t_0 - t \right)^{-\beta - 3}\ .
\ee
Since $R \sim \left(t_0 - t \right)^{-\beta - 1}$ and
$\Box \left(\alpha n/R^{n+1}\right) \sim
\left(t_0 - t \right)^{\left(\beta + 1\right)\left(n+1\right) -2}$,
the terms in the l.h.s of (\ref{CA3}) are always less singular than $T_A$
since
\be
\label{CA5}
 -\beta - 1,\ \left(\beta + 1\right)\left(n+1\right) -2 > -\beta - 3\ .
\ee
This indicates that Eq.(\ref{CA3}) does not allow the singular solution
and the curvature
singularity does not appear.
Therefore, in case $2b/3 + b''\neq 0$, the quantum effects prevent the
singularity appearance.

In the above analysis, $\Box R$ term acts against the singularity.
The $\Box R$ term is generated by a local term $R^2$, which shows that if
one modifies $F(R)$ or $f(R)$ by adding $R^2$ term as
\be
\label{R2A}
F(R) \to F(R) + \gamma R^2\ ,
\ee
with a constant $\gamma$, the curvature singularity could not be generated.
Note also that near to singularity, quantum gravity effects become
dominant. Then, that is quantum gravity which is not constructed yet
should define the universe behaviour near to singularity.

Actually, this observation was made some time ago in ref.\cite{abdalla}
where it was shown that in the presence of $R^2$ term (or higher powers of
curvature) as well as negative
power curvature terms the effective phantom phase is transient and future
singularity never occurs.
This suggests the scenario to avoid the future singularity by the
modification of early-time relevant part of $F(R)$-theory, for instance
by powers of curvature.
Of course, such modification should not destroy the inflationary phase
and its exit as well as perturbations structure. It is also important that
by construction such modification does not violate the local tests
viability of modified gravity.

\section{Realistic model without future singularity}

It has been shown that if $f(R)$ behaves as (\ref{frlv3}) for large curvature
there
could occur the curvature singularity in a finite time.
In the Hu-Sawicki model, the behavior of $f(R)$ in (\ref{frlv3}) generates the
late-time acceleration.

Note, however, the late-time acceleration occurs if $f(R_0)$ becomes
almost constant in the present universe:
\be
\label{JGRG45}
f(R_0)= - 2\tilde R_0\ ,\quad f'(R_0)\sim 0\ .
\ee
Here $R_0$ is the current curvature of the universe and we assume $R_0> \tilde R_0$.
Almost constant $f(R)$ plays the role of the effective small cosmological constant:
$\Lambda_l \sim - f(R_0) = 2\tilde R_0$, which could generate the accelerating expansion
in the present universe.

In the Hu-Sawicki model \cite{HS} the following condition is also satisfied:
\be
\label{JGRG10}
\lim_{R\to 0} f(R) = 0\ .
\ee
The condition shows that there occurs a flat spacetime solution, where
$R=0$.

Then we need not to require the behavior (\ref{frlv3}) when the curvature is really
large $R\gg R_0$. For example, if $f(R)$ behaves as
\be
\label{fr6}
f(R) = - \left(1 - \eta\right) R - 2\eta \Lambda_I + {\cal O}\left(R^{-1}\right)\ ,
\ee
we find
\be
\label{fr7}
F(R) = \eta \left(R - 2\Lambda_I\right) + {\cal O}\left(R^{-1}\right)\ .
\ee
In (\ref{fr6}), $\eta$ is a positive constant.
Eq.(\ref{fr7}) tells that the effective gravitational coupling $\kappa_{\rm eff}$
is given by $\kappa_{\rm eff}^2= \eta \kappa^2$ and the effective cosmological
constant, which could generate the inflation, is given by $\Lambda_I$.
For (\ref{fr7}), the potential $V$ in (\ref{fr4}) is given by
\be
\label{fr8}
V\sim \frac{2\Lambda_I}{\eta}\ .
\ee
Then if $\Lambda_I$ is large enough, the curvature singularity could not
be easily realized.
Hence, if $F(R)$ goes to the Einstein action (the gravitational constant
could be
changed) with the large cosmological constant, the value of the potential for large $R$ becomes
the order of the cosmological constant. Then if the cosmological constant is large enough,
the value of the potential becomes large and the value cannot be easily realized.

An example satisfying the conditions (\ref{JGRG45}), (\ref{JGRG10}), and (\ref{fr6})
could be given by (compare with general viable construction \cite{sergio})
\bea
\label{fr9}
&& f(R) = f_I(R) + f_L(R)\ ,\nn
&& f_I(R) = - \left(1 - \eta\right) R - 2\eta \Lambda_I
+ 2\eta \Lambda_I \e^{(1 - \eta)R/2\eta\Lambda_I
 - (1+\xi)(1 - \eta)^2 R^2 /8\eta^2\Lambda_I^2 }\ ,\quad
f_L(R) = - 2\alpha {R^2}\e^{- c\alpha^2 R^2}\ .
\eea
Here $c$, $\alpha$, and $\xi$ are positive constants satisfying the conditions
$\Lambda_I\gg 1/\alpha$
and $c\sim 1$.
When the curvature has the order of $1/\alpha$, $f_I(R)$ behaves as
$f_I(R) = R\times {\cal O}\left( R/\Lambda_I \right)$ and therefore $f_I(R)$
could be neglected compared with $f_L(R) \sim R$. Then since
\be
\label{fr10}
f'(R) \sim f'_L(R) = - 4\alpha R \left( 1 - c\alpha^2 R^2 \right) \e^{- c\alpha^2 R^2}\ ,
\ee
compared with (\ref{JGRG45}), we find
\be
\label{fr11}
R_0 \sim \frac{1}{\sqrt{c}\alpha}\ ,\quad \tilde R_0 \sim \frac{1}{c\alpha}\e^{-1}\ .
\ee
Since about 70\% of the energy density in the universe is dark energy,
we find $\tilde R_0 \sim 0.7 R_0$. Then one has
\be
\label{fr12}
c \sim (0.7)^2 \e^{-2} \sim 0.07\ .
\ee
We should note that the model (\ref{fr9}) unifies the late acceleration and the inflation
in the early universe like models \cite{Nojiri:2007as,Nojiri:2007cq} and
the future singularity does not appear.

We now check the model (\ref{fr9}) could satisfy the constraints on the correction
to the Newton law. It is convenient to work in the Einstein frame
(\ref{fr4}). The potential $V(\sigma)$
in (\ref{fr4}) gives a mass $m_\sigma$ for the scalar field $\sigma$:
\be
\label{MN3}
m_\sigma^2 \equiv \frac{1}{2}\frac{d^2 V(\sigma)}{d\sigma^2}
=\frac{1}{2}\left\{\frac{A}{F'(A)} - \frac{4F(A)}{\left(F'(A)\right)^2} + \frac{1}{F''(A)}\right\}
\ee
In order that the correction to the Newton law could be small, the mass $m_\sigma$ should be
large enough.
The Newton law has been checked on the earth or in the solar system.
In air on the earth, the scalar curvature could be given by $A=R\sim 10^{-50}\,{\rm eV}^2$.
On the other hand, in the solar system, we find $A=R\sim 10^{-61}\,{\rm eV}^2$.
Since $R_0\sim 10^{-66}\,{\rm eV}^2$, one may assume $\Lambda_I \gg R \gg 1/\alpha$.
Then
\be
\label{fr13}
f_I(R) \sim -\frac{\xi\left( 1 - \eta \right)^2 R^2}{4\eta \Lambda_I}\ ,
\ee
and therefore
\be
\label{fr14}
\left| f_I(R)\right| ,\, \left| f_L(R) \right| \ll R\ ,\quad
\left| f_I'(R)\right| ,\, \left| f_L'(R) \right| \ll 1\ .
\ee
Note that $f_I''(R)$ is negative and
if $\Lambda_I \sim \left(10^{15-18}\,{\rm GeV}\right)^2 = 10^{48-54}\,{\rm eV}^2$,
we have
\be
\label{fr15}
f_I''(R) \sim \left( 10^{48-54}\,{\rm eV}^2 \right)^{-1}\ .
\ee
We also find
\be
\label{fr16}
f_L''(R) \sim - \frac{8}{\sqrt{c} R_0}\left(\frac{R}{R_0}\right)^4 \e^{-R^2/R_0^2}\ ,
\ee
which is negative. Then in the air of the earth, we find
\be
\label{fr17}
f_L''(R) \sim \left( 10^{10^{32}}\,{\rm eV} \right)^{-1}\ ,
\ee
which is extremely small and in the solar system,
\be
\label{fr18}
f_L''(R) \sim \left( 10^{10^{10}}\,{\rm eV} \right)^{-1}\ ,
\ee
which is also extremely small.
Then the mass of scalar field $\sigma$ could be given by
\be
\label{fr19}
m_\sigma^2 \sim - \frac{1}{2F''(R)} \sim - \frac{1}{2f_I''(R)}
\sim 10^{48-54}\,{\rm eV}^2 = \left( 10^{15-18}\,{\rm GeV}\right)^2\ ,
\ee
which is very large and the correction to the Newton law is beoynd of the
observational capacities. Hence, the viable modified gravity which unifies
the inflation with late-time acceleration and which belongs to general
class of ref.\cite{sergio} is free of future singularity.

As it was shown in the arguments around (\ref{f1}), (\ref{f2}) or (\ref{R2A}), instead of
(\ref{fr6}), if we add the term $R^n$ ($0<n\leq 2$), the curvature singularity could be
avoided in the same way as in \cite{abdalla}.

Eq.(\ref{f2}) tells that there could be finite future singularity. We also find that
the Big Rip singularity could be realized in $F(R)\sim R^n$ ($n>2$) theory as
shown in (\ref{rlv13}). This could be confirmed by the trace equation (\ref{Scalaron}).
We now assume (\ref{f1}) with $n>2$ and therefore $F(R) \sim f_n R^n$ with $n>2$ when the
curvature is large. Then the trace equation (\ref{Scalaron}) reduced in the following
form:
\be
\label{Scalaron2}
- 3n\left(\frac{d^2}{dt^2} + 3H\frac{d}{dt}\right)R^{n-1} \sim \left(2 - n\right) R^n \ .
\ee
The explicit solution of (\ref{Scalaron2}) is given by a Big Rip solution:
\be
\label{BR1}
H=\frac{h_\pm^{(n)}}{t_0 - t}\ ,\quad
h_\pm^{(n)}
= \frac{3n^2 -4n + 2 \pm \sqrt{\left( 3n^2 -4n + 2 \right)^2 + 8(n-2)(n-1)n(2n-1)}}
{4(n-2)}\ .
\ee
Since $3n^2 - 4n + 2 = 3(n-2)^2 + 8(n-2) + 4$, $h_+^{(n)}$ is surely real and positive.
Then there could appear finite future Big Rip singularity in $F(R)\sim R^n$ ($n>2$) theory.
However, considering several such terms with different powers of curvature
may often lead to stable de Sitter solution which prevents the evolution
to singularity.

\section{Discussion}

In summary, we studied the future evolution in the models of
$F(R)$-gravity consistent with local tests and unifying the universe
expansion history. Working in both (Einstein and Jordan) frames we
demonstrated how Big Rip or type II,III and IV future singularities
appear. Using the reconstruction method \cite{NOr} it is shown which
 models may lead to any specific type future singularity. It is remarkable
that whatever type singularity may occur in Jordan frame, in the
corresponding Einstein frame the only IV type singularity appears.
It is also interesting that as in the scalar-tensor theory, the appearing
singularity does not lead to the violation of local tests when modified
gravity is consistent with them. Nevertheless, some indications to
possible future singularity may be expected to be found in the current
observational bounds\cite{Nojiri:2007cq,yurov}.

Even if $F(R)$-gravity under investigation develops the future
singularity,
there is always the way to modify its structure by terms relevant at the
early universe. Such terms, like in the model \cite{abdalla}, prevent the
development of the singularity.
The explicit example of realistic $F(R)$-gravity which does not lead to
future singularity even classically is proposed.
>From another point, the classical description breaks down near to
singularity. The account of quantum effects becomes necessary in this
situation. Taking into account the quantum effects of conformal fields,
we show that such quantum effects naturally act against the singularity
appearence. Of course, in order to understand how realistic is our quantum
description of future singularity one should take into account quantum
gravity.

As the final remark, let us note that other modified gravities
(Gauss-Bonnet gravity, string-inspired gravity, theory with inhomogeneous
equation of state) \cite{review} also naturally describe the
quintessence/phantom dominated universe which may evolve to future
singularity. It would be of interest to study the structure of future
singularities in such models too.

\section*{Acknowledgments}

We are grateful to M. Sami for collaboration at the early stage of this work.
The work by S.D.O. is
supported in part by MEC (Spain) projects FIS2006-02842 and
PIE2007-50/023 and RFBR grant 06-01-00609 and LRSS project N.2553.2008.2.
The work by S.N. is supported in part by the Ministry of Education,
Science, Sports and Culture of Japan under grant no.18549001.

\end{document}